%Paper: hep-th/9506077
%From: Nathan Seiberg <seiberg@physics.rutgers.edu>
%Date: Sun, 11 Jun 1995 19:42:52 -0400

%uses harvmac.tex
\input harvmac

\font\zfont = cmss10 %scaled \magstep1

\def\bigone{\hbox{1\kern -.23em {\rm l}}}
\def\ZZ{\hbox{\zfont Z\kern-.4emZ}}

\noblackbox

\def\np#1#2#3{Nucl. Phys. B{#1} (#2) #3}
\def\pl#1#2#3{Phys. Lett. {#1}B (#2) #3}
\def\prl#1#2#3{Phys. Rev. Lett. {#1} (#2) #3}
\def\physrev#1#2#3{Phys. Rev. {D#1} (#2) #3}

\def\prep#1#2#3{Phys. Rep. {#1} (#2) #3}

\def\cmp#1#2#3{Comm. Math. Phys. {#1} (#2) #3}

\Title{hep-th/9506077, RU-95-37, IASSNS-HEP-95/46}
{\vbox{\centerline{The Power of Duality -- }
\bigskip
\baselineskip=16pt
\centerline{Exact Results in 4D SUSY Field Theory}}}
\bigskip
\centerline{N. Seiberg
\footnote{}{To appear in the Proc.\ of PASCOS 95 and in the Proc.\ of
the Oskar Klein Lectures.}
}
\vglue .5cm
\centerline{Department of Physics and Astronomy}
\centerline{Rutgers University}
\centerline{Piscataway, NJ 08855-0849, USA}
\vglue .3cm
\centerline{and}
\vglue .3cm
\centerline{Institute for Advanced Study}
\centerline{Princeton, NJ 08540, USA}

\bigskip\bigskip

\noindent
Recently the vacuum structure of a large class of four dimensional
(supersymmetric) quantum field theories was determined exactly.  These
theories exhibit a wide range of interesting new physical phenomena.
One of the main new insights is the role of ``electric-magnetic
duality.''  In its simplest form it describes the long distance behavior
of some strongly coupled, and hence complicated, ``electric theories''
in terms of weakly coupled ``magnetic theories.''  This understanding
sheds new light on confinement and the Higgs mechanism and uncovers new
phases of four dimensional gauge theories.  We review these developments
and speculate on the outlook.

\Date{6/95}
%\draftmode

\newsec{Introduction}

Exact solutions play a crucial role in physics.  It is often the case
that a simple model exhibits the same phenomena which are also present
in more complicated examples.  The exact solution of the simple model
then teaches us about more generic situations. For example, the exact
solutions of the harmonic oscillator and of the hydrogen atom
demonstrated many of the crucial aspects of quantum mechanics.
Similarly, the Ising model was a useful laboratory in the study of
statistical mechanics and quantum field theory.  Many other exactly
solvable two dimensional field theories like the Schwinger model and
others have led to the understanding of many mechanisms in quantum field
theory, which are also present in four dimensions.

The main point of this talk is to show how four dimensional
supersymmetric quantum field theories can play a similar role as
laboratories and testing grounds for ideas in more generic quantum field
theories. This follows {}from the fact that these theories are more
tractable than ordinary, non-supersymmetric theories, and many of their
observables can be computed exactly.  Nevertheless, it turns out that
these theories exhibit explicit examples of various phenomena in quantum
field theory.  Some of them had been suggested before without an
explicit realization and others are completely new.  (For a brief review
summarizing the understanding as of a year ago, see
\ref\pwer{N. Seiberg, The Power of Holomorphy -- Exact Results in 4D
SUSY Field Theories.  To appear in the Proc. of PASCOS 94.
hep-th/9408013, RU-94-64, IASSNS-HEP-94/57}.)

Before continuing we would like to mention that supersymmetric four
dimensional field theories also have two other applications:

\item{1.} Many physicists expect supersymmetry to be present in
Nature, the reason being that it appears to be the leading candidate for
solving the gauge hierarchy problem.  If this is indeed the case, it is
likely to be discovered experimentally in the next round of
accelerators.  Independent of the hierarchy problem, supersymmetry plays
an important role in string theory as an interesting extension of our
ideas about space and time.  If Nature is indeed supersymmetric,
understanding the dynamics of supersymmetric theories will have direct
experimental applications.  In particular, we will have to understand
how supersymmetry is broken, i.e.\ why Nature is not exactly
supersymmetric.

\item{2.} Witten discovered an interesting relation between
supersymmetric field theories and four dimensional topology
\ref\tft{E. Witten, \cmp{117}{1988}{353}.}.
Using this relation, the exact solutions lead to a simplification of
certain topological field theories and with that to advances in
topology
\ref\monop{E. Witten, hep-th/9411102, Math. Res. Lett. 1 (1994) 769.}.

Even though these applications are important, here we will focus on the
dynamical issues and will not discuss them.

The main insight that the study of these theories has taught us so far
is the role of electric-magnetic duality in strongly coupled non-Abelian
gauge theories\foot{The role of electric-magnetic duality in four
dimensional quantum field theory was first suggested by Montonen and
Olive
\ref\mo{C. Montonen and D. Olive, \pl {72}{1977}{117}; P. Goddard,
J. Nuyts and D. Olive, \np{125}{1977}{1}.}.
Then, it became clear that the simplest version of their proposal is
true only in $N=4$ supersymmetric field theories
\ref\dualnf{H. Osborn, \pl{83}{1979}{321}; A. Sen, hep-th/9402032,
\pl{329}{1994}{217}; C. Vafa and E. Witten, hep-th/9408074,
\np{432}{1994}{3}.}
and in certain $N=2$ supersymmetric theories
\ref\swii{N. Seiberg and E. Witten, hep-th/9408099,
\np{431}{1994}{484}.}.
Here we will discuss the extension of these ideas to $N=1$ theories
\ref\sem{N. Seiberg, hep-th/9411149, \np {435}{1995}{129}.}.}.
Therefore, we will start our discussion in the next section by reviewing
the duality in Abelian gauge theories, i.e.\ in electrodynamics.  In
section 3 we will present supersymmetric field theories and will outline
how they are solved.  This discussion will be rather heuristic.  In
section 4 we will summarize the results and will present the duality in
non-Abelian theories.  Finally in section 5 we will present our
conclusions and a speculative outlook.

\newsec{Duality in electrodynamics}

\subsec{The Coulomb phase}

The simplest phase of electrodynamics is the Coulomb phase.  It is
characterized by massless photons which mediate a long range $1\over R$
potential between external sources.  In the absence of sources the
relevant equations are
Maxwell's equations in the vacuum
$$\eqalign{
& \nabla \cdot E =0 \cr
& \nabla \times B - {\partial \over \partial t} E =0 \cr
& \nabla \cdot B =0 \cr
& \nabla \times E + {\partial \over \partial t} B =0 \cr} $$
(we have set the speed of light $c=1$).
Clearly, they are invariant under the duality transformation
$$\eqalign{ &E ~~\longrightarrow ~~ B
\cr &B ~~\longrightarrow ~~ -E \cr}$$
which exchanges electric and magnetic fields.

If charged particles are added to the equations, the duality symmetry
will be preserved only if both electric charges and magnetic monopoles
are present.  However, in Nature we see electric charges but no magnetic
monopole has been observed yet. This fact ruins the duality symmetry.
We usually also ruin the symmetry by solving two of the equations by
introducing the vector potential.  Then, these equations are referred to
as the Bianchi identities while the other two equations are the
equations of motion.  Had there also been magnetic monopoles, this would
have been impossible.

Dirac was the first to study the possible existence of magnetic
monopoles in the quantum theory.  He derived the famous Dirac
quantization condition which relates the electric charge $e$ and the
magnetic charge $g$:
$$e g=2\pi$$
(we have set Planck's constant $\hbar=1$).
This relation has many important consequences.  One of them is that
since duality exchanges electric and magnetic fields, it also exchanges
$$ e ~~\longleftrightarrow ~~ g.$$
Since the product $eg=2\pi$ is fixed, it relates weak coupling ($e\ll
1$) to strong coupling ($g \gg 1$).  Therefore, even if we could perform
such duality transformations in quantum electrodynamics, they would
not be useful.  Electrodynamics is weakly coupled because $e$ is small.
Expressing it in terms of ``magnetic variables'' will make it strongly
coupled and therefore complicated.

However, one might hope that in other theories like QCD, such a duality
transformation exists.  If it does, it will map the underlying
``electric'' degrees of freedom of QCD, which are strongly coupled, to
weakly coupled ``magnetic'' degrees of freedom.  We would then have a
weakly coupled, and therefore easily understandable, effective
description of QCD.

\subsec{The Higgs phase}

When charged matter particles are present, electrodynamics can be in
another phase -- the superconducting or the Higgs phase.  It is
characterized by the condensation of a charged field $\phi$
$$\langle \phi \rangle \not=0.$$
This condensation creates a gap in the spectrum by making the photon
massive.  This phenomenon was first described in the context of
superconductivity, where $\phi$ is the Cooper pair.  It has since
appeared in different systems including the weak interactions of
particle physics where $\phi$ is the Higgs field.

The condensation of $\phi$ makes electric currents superconducting.  Its
effect on magnetic fields is known as the Meissner effect.  Magnetic
fields cannot penetrate the superconductor except in thin flux tubes.
Therefore, when two magnetic monopoles (e.g.\ the ends of a long magnet)
are inserted in a superconductor, the flux lines are not spread.
Instead, a thin flux tube is formed between them.  The energy stored in
the flux tube is linear in its length and therefore the potential
between two external magnetic monopoles is linear (as opposed to the $1
\over R$ potential outside the superconductor).  Such a linear
potential is known as a {\it confining} potential.

Mandelstam and 'tHooft considered the dual of this phenomenon: if
instead of electric charges, magnetic monopoles condense, then magnetic
currents are superconducting while electric charges are confined.
Therefore, confinement is the dual of the Higgs mechanism.  They
suggested that confinement in QCD can be understood in a similar way by
the condensation of color magnetic monopoles.

To summarize, we see that duality exchanges weak coupling and strong
coupling.  Therefore, it exchanges a description of the theory with
small quantum fluctuations with a description with large quantum
fluctuations.  Similarly, it exchanges the weakly coupled phenomenon of
the Higgs mechanism with the strong coupling phenomenon of confinement.

Such a transformation between variables which fluctuate rapidly and
variables which are almost fixed is similar to a Fourier transform.
When a coordinate is localized, its conjugate momentum fluctuates
rapidly and vice versa.  A duality transformation is like a
Fourier transform between electric and magnetic variables.

It should be stressed, however, that except in simple cases (like
electrodynamics without charges and some examples in two dimensional
field theory) an explicit duality transformation is not known.  It is
not even known whether such a transformation exists at all.  As we will
show below, at least in supersymmetric theories such a transformation
does exist (even though we do not have an explicit description of it).

\newsec{The Dynamics of Supersymmetric Field Theories}

In supersymmetric theories the elementary particles are in
representations of supersymmetry.  Every gauge boson, a gluon, is
accompanied by a fermion, a gluino, and every matter fermion, a quark,
is accompanied by a scalar, a squark.  The theory is specified by a
choice of a gauge group, which determines the coupling between the
gluons, and a matter representation, which determines the coupling
between the quarks and the gluons.  For example, in QCD the gauge
group is $SU(N_c)$ ($N_c$ is the number of colors) and the matter
representation is $N_f \times ({\bf N_c} + {\bf \overline N_c})$ ($N_f$
is the number of flavors).

An important object in these theories is the superpotential, $W(q)$.  It
is a holomorphic (independent of $\bar q$, the complex conjugate of $q$)
gauge invariant function of the squarks, $q$. It determines many of the
coupling constants and interactions in the theory including the Yukawa
couplings of the quarks and the squarks and the scalar potential,
$V(q,\bar q)$.  For example, a mass term for the quarks appears as a
quadratic term in the superpotential.

The analysis of the classical theory starts by studying the minima of
the scalar potential $V(q,\bar q)$.  It is often the case that the
potential has many different degenerate minima.  Then, the classical
theory has many inequivalent ground states.  Such a degeneracy between
states which are not related by a symmetry is known as an ``accidental
degeneracy.''  Typically in field theory such an accidental degeneracy
is lifted by quantum corrections.  However, in supersymmetric theories,
the degeneracy often persists in the quantum theory.  Therefore, the
quantum theory has a space of inequivalent vacua which can be labeled by
the expectation values of the squarks $\langle q\rangle $.

This situation of many different ground states in the quantum theory is
reminiscent of the situation when a symmetry is spontaneously broken.
However, it should be stressed, that unlike that case, where the
different ground states are related by a symmetry, here they are
inequivalent.  Physical observables vary {}from one vacuum to the other
-- they are functions of $\langle q\rangle $.

Our problem is to solve these theories not only as a function of all the
parameters (like quark masses) but, for every value of the parameters,
also as a function of the ground state $\langle q\rangle $.

As in Landau-Ginzburg theory, the best way to organize the information
is in a low energy effective Lagrangian, $L_{\rm eff}$.  It includes
only the low lying modes and describes their interactions.  This
effective theory is also specified by a gauge group and a matter
representation.  Since the theory is supersymmetric\foot{We limit
ourselves to theories where supersymmetry is not spontaneously broken.},
the interactions of the light particles are characterized by a
superpotential, $W_{\rm eff}$.

The key observation is that this effective superpotential can often be
determined exactly by imposing the following constraints
\ref\nonren{N. Seiberg, hep-ph/9309335, \pl{318}{1993}{469}.}:

\item{1.}  In various limits of the parameter space or the space of
ground states the theory is weakly coupled.  In these limits $W_{\rm
eff}$ can be determined approximately by weak coupling techniques
(examples are the instanton calculations of
\nref\ads{I. Affleck, M. Dine and N. Seiberg, \np{241}{1984}{493};
\np{256}{1985}{557}.}%
\nref\natint{N. Seiberg, \pl{206}{1988}{75}.}%
\nref\finnpou{D. Finnell and P. Pouliot, RU-95-14, SLAC-PUB-95-6768,
hep-th/9503115.}%
\refs{\ads - \finnpou}).

\item{2.} $W_{\rm eff}$ must respect all the symmetries in the problem.
It is important to use also the constraints following {}from symmetries
which are explicitly broken by the various coupling constants.  Such
symmetries lead to selection rules.

\item{3.} $W_{\rm eff}$ is holomorphic in the light fields
and the parameters.  This constraint is the crucial one which makes
supersymmetric theories different {}from non-supersymmetric ones.  The
use of holomorphy in \nonren\ generalized previous related ideas
\nref\witholo{E. Witten, \np{268}{1986}{79}.}%
\nref\dinese{M. Dine and N. Seiberg, \prl{57}{1986}{2625}.}%
\nref\sv{M.A. Shifman and A. I. Vainshtein, \np{277}{1986}{456};
\np{359}{1991}{571}.}%
\nref\cern{D. Amati, K. Konishi, Y. Meurice, G.C. Rossi and G.
Veneziano, \prep{162}{1988}{169} and references therein.}%
\nref\polch{J. Polchinski and N. Seiberg, (1988) unpublished.}%
\refs{\witholo-\polch}.

The superpotential determined in this way teaches us about the light
particles and their interactions.  This information determines the phase
structure of the theory and the mechanisms for phase transitions.

By repeating this process for different theories, i.e.\ different gauge
groups and matter representations, a rich spectrum of phenomena has been
found.  In the next section we will describe some of them.

\newsec{Results -- duality in non-Abelian theories}

\nref\nsvz{V.A. Novikov, M.A. Shifman, A. I.  Vainshtein and V. I.
Zakharov, \np{223}{1983}{445}; \np{229}{1983}{381};
\np{260}{1985}{157}.}%
\nref\nati{N. Seiberg, hep-th/9402044, \physrev{49}{1994}{6857}.}%

\subsec{$SU(N_c)$ with $N_f \times ({\bf N_c} + {\bf \overline
N_c})$}

Here the physical phenomena depend crucially on the number of massless
quarks \refs{\ads, \nsvz, \cern, \nati, \sem}:

\bigskip\noindent
I. $N_f \ge 3N_c$

In this range the theory is not asymptotically free.  This means that,
because of screening, the coupling constant becomes smaller at large
distances.  Therefore, the spectrum of the theory at large distance can
be read off {}from the Lagrangian -- it consists of the elementary
quarks and gluons.  The long distance behavior of the potential between
external electric sources is of the form
$$V \sim {1 \over R \log R}.$$
The logarithm in the denominator shows that the interactions between
the particles are weaker than in Coulomb theory.  Therefore, we refer
to this phase of the theory as a free electric phase.

We should add here that, strictly speaking, such a theory is not well
defined as an interacting quantum field theory. However, it can be a
consistent description of the low energy limit of another theory.

\bigskip\noindent
II. ${3 \over 2}N_c < N_f < 3N_c$

In this range the theory is asymptotically free.  This means that at
short distance the coupling constant is small and it becomes larger at
longer distances.  However, in this regime \refs{\nati, \sem} rather
than growing to infinity, it reaches a finite value -- a fixed point of
the renormalization group.

Therefore, this is a non-trivial four dimensional conformal field
theory.  The elementary quarks and gluons are not confined there but
appear as interacting massless particles.  The potential between external
electric sources behaves as
$$V \sim {1 \over R}$$
and therefore we refer to this phase of the theory as the non-Abelian
Coulomb phase.

It turns out that there is an equivalent, ``magnetic,'' description of
the physics at this point \sem.  It is based on the gauge group
$SU(N_f-N_c)$, with $N_f$ flavors of quarks and some gauge invariant
fields.  We will refer to this gauge group as the magnetic gauge group
and to its quarks as magnetic quarks.  This theory is also in a
non-Abelian Coulomb phase because ${3 \over 2}(N_f-N_c) < N_f <
3(N_f-N_c)$.  The surprising fact is that its large distance behavior is
identical to the large distance behavior of the original, ``electric,''
$SU(N_c)$ theory.  Note that the two theories have different gauge
groups and different numbers of interacting particles.  Nevertheless,
they describe the same fixed point.  In other words, there is no
experimental way to determine whether the $1 \over R$ potential between
external sources is mediated by the interacting electric or the
interacting magnetic variables.  Such a phenomenon of two different
Lagrangians describing the same long distance physics is common in two
dimensions and is known there as quantum equivalence.  These four
dimensional examples generalize the duality \mo\ in finite $N=4$
supersymmetric theories \dualnf\ and in finite $N=2$ theories \swii\ to
asymptotically free $N=1$ theories.

As $N_f$ is reduced (e.g.\ by giving masses to some quarks and
decoupling them) the electric theory becomes stronger -- the fixed point
of the renormalization group occurs at larger values of the coupling.
Correspondingly, the magnetic theory becomes weaker.  This can be
understood by noting that as $N_f$ is reduced, the magnetic gauge group
becomes smaller.  This happens by the Higgs mechanism in the magnetic
theory \sem.

\bigskip\noindent
III. $N_c+2 \le N_f \le {3 \over 2} N_c$

In this range the electric theory is very strongly coupled.  However,
since $3(N_f-N_c) \le N_f$, the equivalent magnetic description based on
the gauge group $SU(N_f-N_c)$ is not asymptotically free and it is
weakly coupled at large distances.  Therefore, the low energy spectrum
of the theory consists of the particles in the dual magnetic Lagrangian
\sem.  These magnetic massless states are composites of the elementary
electric degrees of freedom.  The massless composite gauge bosons
exhibit gauge invariance which is not visible in the underlying
electric description.  The theory generates new gauge invariance!  We
will return to this phenomenon in the conclusions.

This understanding allows us to determine the long distance behavior of
the potential between magnetic sources:
$$V \sim {1 \over R \log R}.$$
Since the magnetic variables are free at long distance, we refer to this
phase as a free magnetic phase.

\bigskip\noindent
IV. $N_f=N_c+1, N_c$

As we continue to decouple quarks by giving them masses and thus
reducing $N_f$, the magnetic gauge group is Higgsed more and more.
Eventually it is completely broken and there are no massless gauge
bosons.  This complete Higgsing of the magnetic theory can be
interpreted as complete confinement of the electric variables.  We thus
see an explicit realization of the ideas of Mandelstam and 'tHooft about
confinement.

For $N_f=N_c+1$ this confinement is not accompanied by chiral symmetry
breaking, while for $N_f=N_c$ chiral symmetry is also broken \nati.  In
the electric language we describe the spectrum in terms of gauge
invariant fields.  These include massless mesons and baryons.  In the
magnetic language these are elementary fields. The idea that some of the
composites in QCD, in particular the baryons, can be thought of as
solitons was suggested in
\ref\barsol{T.H.R. Skyrme, Proc.Roy.Soc. A260 (1961) 127; E. Witten, \np
{160}{1979}{57}; \np{223}{1983}{422}; \np {223}{1983}{433}.}.
Here we see an explicit realization of a related idea -- the baryons
are magnetic monopoles composed of the elementary quarks and gluons.

\bigskip\noindent
V. $N_f <N_c$

In this range the theory of massless quarks has no ground state
\refs{\ads, \cern}.

\subsec{$SO(N_c)$ with $N_f \times {\bf N_c}$}

In the $SU(N_c)$ theories there is no invariant distinction between
Higgs and confinement
\ref\higgscon{T. Banks, E. Rabinovici, \np{160}{1979}{349};
E. Fradkin and S. Shenker, \physrev{19}{1979}{3682}.}.
This is not the case in theories based on $SO(N_c)$ with $N_f
\times {\bf N_c}$ and therefore, they lead to a clearer picture of the
dynamics.  In particular, here the transition {}from the Higgs phase to
the Confining phase occurs with a well defined phase transition.

\nref\ils{K. Intriligator, R.G. Leigh and N. Seiberg, hep-th/9403198,
\physrev{50}{1994}{1092}; K. Intriligator, hep-th/9407106,
\pl{336}{1994}{409}.}%
\nref\swi{N. Seiberg and E. Witten, hep-th/9407087,
\np{426}{1994}{19}.}%
\nref\intse{K. Intriligator and N. Seiberg, hep-th/9408155,
\np{431}{1994}{551}.}%
\nref\intseso{K. Intriligator and N. Seiberg, RU--95--3,
hep-th/9503179; RU-95-40, IASSNS-HEP-95/48, to appear in the Proc. of
Strings 95.}%

Many of the results in these $SO(N_c)$ theories
\refs{\ads, \ils-\intse, \sem, \intseso}
are similar to the results in $SU(N_c)$ showing
that these phenomena are generic.  Here the duality map is
$$SO(N_c) ~ {\rm with} ~ N_f \times {\bf N_c}
\longleftrightarrow SO(N_f-N_c+4) ~ {\rm with} ~ N_f \times {\bf
(N_f-N_c+4)}$$

Let us consider three special cases of the duality:

\item{1.} For $N_c=2$, $N_f=0$ the map is
$$SO(2) \cong U(1)\longleftrightarrow SO(2) \cong U(1) $$
which is the ordinary duality of electrodynamics.  Therefore, our
duality is compatible with and generalizes this duality.

\item{2.} For $N_c=3$, $N_f=1$ the map is:
$$SO(3)~ {\rm with}~ {\bf 3}\longleftrightarrow SO(2) \cong  U(1) ~{\rm
with}~~ {\bf 2}. $$
This theory was first analyzed in \swi.  Here it is possible to
understand the duality in more detail than in the more general case and
in particular to identify the matter fields in the magnetic theory as
magnetic monopoles.  This theory has $N=2$ supersymmetry but this did
not play a role in our discussion.  However, using the extra
supersymmetry one can derive more exact results about the massive
spectrum of the theory \swi.

\item{3.} For $N_f=N_c-2$ the map is
$$SO(N_c)~ {\rm with}~  (N_c-2) \times {\bf N_c}
\longleftrightarrow SO(2) \cong  U(1) ~{\rm with}~~(N_c-2) \times
{\bf 2}$$
which generalizes the previous example to situations without $N=2$
supersymmetry.

\noindent
These three examples are simple because the magnetic theory is Abelian.
However, in general the magnetic $ SO(N_f-N_c+4)$ gauge group is
non-Abelian.

These $SO(N_c)$ theories also exhibit many new phenomena, which are not
present in the $SU(N_c)$ examples.  The most dramatic of them is oblique
confinement
\nref\thooft{G. 'tHooft, \np{190}{1981}{455}.}%
\nref\cardyrabin{J. Cardy and E. Rabinovici, \np{205}{1982}{1};
J. Cardy, \np{205}{1982}{17}.}%
\refs{\thooft,\cardyrabin}, driven by the condensation of dyons
(particles with both electric and magnetic charges).
This phenomenon is best described by another equivalent theory -- a
dyonic theory.  Therefore, these theories exhibit
electric-magnetic-dyonic triality \intseso.

\newsec{Conclusions}

To conclude, supersymmetric field theories are tractable and many of
their observables can be computed exactly\foot{Although we did not
discuss them here, we would like to point out that many other
examples were studied
\nref\mos{A.Yu. Morozov, M.A. Olshansetsky and M.A. Shifman,
\np{304}{1988}{291}.}%
\nref\iss{K. Intriligator, N. Seiberg and S. Shenker, hep-ph/9410203,
\pl {342}{1995}{152}.}%
\nref\sunnt{A. Klemm, W. Lerche, S. Theisen and S. Yankielowicz,
hep-th/9411048, \pl{344}{1995}{169}; hep-th/9412158.}%
\nref\arfa{P. Argyres and A. Faraggi, hep-th/9411057.}%
\nref\aharony{O. Aharony, hep-th/9502013, TAUP-2232-95.}%
\nref\kutasov{D. Kutasov, hep-th/9503086, EFI-95-11.}%
\nref\rlms{R. Leigh and M. Strassler, RU-95-2, hep-th/9503121.}%
\nref\dansun{U. Danielsson and B. Sundborg, USITP-95-06, UUITP-4/95,
hep-th/9504102.}%
\nref\doush{M.R. Douglas and S.H. Shenker, RU-95-12, RU-95-12,
hep-th/9503163.}%
\nref\jerus{S. Elitzur, A Forge, A. Giveon and E. Rabinovici, RI-4-95
hep-th/9504080.}%
\nref\asy{O. Aharony, J. Sonnenschein and S. Yankielowicz,
TAUP--2246--95, CERN-TH/95--91, hep-th/9504113.}%
\nref\schkut{D. Kutasov and A. Schwimmer, EFI--95--20, WIS/4/95,
hep-th/9505004.}%
\nref\intpou{K. Intriligator and P. Pouliot, RU-95-23, hep-th/9505006.}%
\nref\intdual{K. Intriligator, RU--95--27, hep-th/9505051.}%
\nref\ad{P.C. Argyres and M.R. Douglas, RU-95-31, hep-th/9505062.}%
\nref\berkooz{M. Berkooz, RU-95-29, hep-th/9505067.}%
\nref\rlmsspso{R. Leigh and M. Strassler, hep-th/9505088, RU-95-30.}%
\nref\ntwomatti{A. Hanany and Y. Oz, TAUP-2248-95, WIS-95/19,
hep-th/9505075.}%
\nref\ntwomattii{P.C. Argyres, M.R. Plesser and A. Shapere,
IASSNS-HEP-95/32, UK-HEP/95-06, hep-th/9505100.}%
\nref\ilst{K. Intriligator, R. Leigh and M. Strassler, RU-95-38, to
appear.}%
\refs{\ads, \cern, \mos, \ils, \swii, \intse, \sem, \iss-\ilst}
exhibiting many new interesting phenomena.}.
The main dynamical lesson we
learn is the role of electric-magnetic duality in non-Abelian gauge
theories in four dimensions.  This duality generalizes the duality in
Maxwell's theory and in $N=4$ \mo\ and certain $N=2$
supersymmetric theories \swii.

The magnetic degrees of freedom are related to the underlying electric
degrees of freedom in a complicated (non-local) way.  They are the
effective degrees of freedom useful for describing the long distance
behavior of the theory.  These variables give a weak coupling
description of strong coupling phenomena such as confinement.

Our analysis led us to find new phases of non-Abelian gauge theories,
like the non-Abelian Coulomb phase with its quantum equivalence and the
free magnetic phase with its massless composite gauge bosons.

\bigskip

\centerline{\it Outlook}

We would like to end by suggesting some future directions for research
and speculate about them.

\item{1.}
The exploration of supersymmetric models is far {}from complete.  There
are many models whose dynamics are not yet understood.  It is likely
that there are new phenomena in quantum field theory which can be
uncovered here.

\item{2.} Of particular importance are chiral theories whose matter
content is not in a real representation of the gauge group.  Only a few
of these have been analyzed \refs{\ads, \cern, \iss}.  These theories
are also interesting as they can lead to dynamical supersymmetry
breaking.  Finding a nice model of dynamical supersymmetry breaking
which is phenomenologically acceptable is an important challenge.  For
some recent work in this direction see
\ref\dinenelson{M. Dine and A.E. Nelson and Y. Shirman, hep-ph/9408384,
\physrev{51}{1995}{1362}.}.

\item{3.}
An important question is to what extent these results are specific to
supersymmetric theories.  It would be very interesting to extend at
least some of these ideas to non-supersymmetric theories and to find the
various phases and the mechanisms for the phase transitions without
supersymmetry.  One way such a study can proceed is by perturbing a
supersymmetric theory whose solution is known by soft breaking terms.
When these terms are small they do not affect the dynamics significantly
and the non-supersymmetric perturbed theory is qualitatively similar to
the unperturbed theory (for a recent attempt in this direction see
\ref\yale{N. Evans, S.D.H. Hsu, M. Schwetz, hep-th/9503186,
YCTP-P8-95.}).
This makes it clear that the phenomena we found can be present in
non-supersymmetric theories as well.  Alternatively, one can start with
ordinary non-supersymmetric QCD with a large enough number of flavors
$N_f$, where a non-trivial fixed point of the renormalization group
exists
\ref\banksz{T. Banks and A. Zaks, \np{196}{1982}{189}.}
and the theory is in a non-Abelian Coulomb phase.  It is not yet known
for which values of $N_f$ this phase exists.  It is possible that there
is a dual magnetic description of this phase and perhaps even a
non-Abelian free magnetic phase exists for some values of $N_f$.  One
might be tempted to speculate that perhaps the confinement of ordinary
QCD can be described as a Higgs phenomenon in these variables.  Then,
perhaps some of the massive particles in the spectrum of QCD (like the
$\rho$ meson or the $A_1$) can be identified as ``Higgsed magnetic
gluons.''

\item{4.}
The duality points at a big gap in our current understanding of gauge
theories.  We do not have an explicit transformation relating the
underlying electric degrees of freedom to their magnetic counterparts.
Finding such an explicit transformation will be extremely interesting.
If such a transformation does not exist within the standard framework of
local quantum field theory, perhaps a reformulation of quantum field
theory will be needed.

\item{5.}
The duality reflects new gauge invariance, which is not obvious in the
fundamental description of the theory.  In hindsight this is not
surprising because gauge symmetry is not a symmetry.  It is merely a
redundancy in the description.  Perhaps we should conclude that gauge
symmetries might not be fundamental!  They might only appear as long
distance artifacts of our description of the theory.  If so, perhaps
some of the gauge symmetries of the standard model or even general
relativity are similarly long distance artifacts.  Then, the
corresponding gauge particles are the ``magnetic'' degrees of freedom of
more elementary ``electric'' variables.  In order not to violate the
theorem of
\ref\ww{S. Weinberg and E. Witten, \pl{96}{1980}{59}.},
the underlying theory cannot have these symmetries as global symmetries.
In particular, this can be the case for gravity only if the underlying
theory is topological.

\item{6.}
Recently, there has been enormous interest in dualities and
non-perturbative effects in string theory
\ref\dualst{At the current rate of progress any list of references will
be enormous and will immediately become outdated.}.
Some of the phenomena found in string theory are similar to and
generalize those found in field theory.  It would be nice to use some of
the techniques which turned out to be useful in field theory in string
theory as well.

Finally we should note that there are obvious relations between these
directions.  For example, progress in understanding the origin of the
duality can help the study of supersymmetric chiral theories and lead to
phenomenological theories of supersymmetry breaking.

\bigskip
\centerline{{\bf Acknowledgments}}

We would like to thank T. Banks, D. Kutasov, R. Leigh, M.R.  Plesser, P.
Pouliot, S.  Shenker, M. Strassler and especially K.  Intriligator and
E. Witten for many helpful discussions.  This work was supported in part
by DOE grant \#DE-FG05-90ER40559.

\listrefs
\end